\documentclass[aps,pra,superscriptaddress,12pt,tightenlines,nofootinbib]{revtex4}
\usepackage{amsmath,amsthm,graphicx,amssymb,verbatim,listings}
\usepackage[T1]{fontenc}     % needed for the guillemets
\usepackage{lmodern}         % needed to fix suboptimal font rendering
\usepackage[ pdftex, plainpages = false, pdfpagelabels,
                 pdfpagelayout = useoutlines,
                 bookmarks,
                 bookmarksopen = true,
                 bookmarksnumbered = true,
                 breaklinks = true,
                 linktocpage=all,
                 pagebackref=false,
                 colorlinks = true,
                 linkcolor = BrickRed,
                 urlcolor  = blue,
                 citecolor = BrickRed,
                 anchorcolor = green,
                 hyperindex = true,
                 hyperfigures
                 ]{hyperref}
\usepackage[usenames, dvipsnames]{xcolor}
\usepackage{xifthen} % provides \isempty test

\newcommand{\ket}[1]{{\ensuremath{\left| #1 \right\rangle}}}

\newcommand{\arxiv}[2][]{\ifthenelse{\isempty{#1}}{\href{http://arxiv.org/abs/#2}{{\tt arXiv:\allowbreak{}#2}}} {\href{http://arxiv.org/abs/#2}{{\tt arXiv:\allowbreak{}#2 [#1]}}}}

\newcommand{\booktitle}{\textsl}
\newcommand{\hrefdoi}[2]{\href{https://dx.doi.org/#1}{#2}}

\begin{document}

\title{The De-Relationalizing of Relational Quantum Mechanics}

\author{Blake C.\ Stacey}
\affiliation{Physics Department, University of Massachusetts Boston}

\date{\today}

\begin{abstract}
A recent phase transition in the relational interpretation of quantum
mechanics (RQM) is situated in its historical context, and the novelty
of the post-transition viewpoint is questioned.
\end{abstract}

\maketitle

Relational Quantum Mechanics (RQM) is an approach to interpreting
quantum theory first proposed by Rovelli in the Before
Times~\cite{Rovelli:1996}. Recently, RQM has been critiqued by several
authors~\cite{Mucino:2021, Pienaar:2021, Brukner:2021, Mucino:2021b,
  Pienaar:2021b, Stacey:2021}. Perhaps in response to some of these
critiques, Adlam and Rovelli published a significant revision to RQM
which, in so many words, backs away from the relationalism that had
characterized the interpretation~\cite{Adlam:2022}. The change can be
illustrated by a thought-experiment in the vein of Wigner's Friend, or
the ``observer observed''. Bob stands outside a room, which he
believes to contain Alice and a system, e.g., a qubit. We can
contemplate two different kinds of consistency conditions regarding
this scenario. Suppose that Bob can measure either Alice, the qubit,
or both in succession. Bob might ask Alice, ``What did you get when
you measured the qubit in the $\sigma_z$ basis?'' He might then expect
that if he measures the qubit in the $\sigma_z$ basis himself, he will
get the same answer that he heard from Alice. This is a
\emph{first-person} consistency condition:\ It is phrased entirely in
terms of what Bob experiences and expects. We can also imagine a
\emph{third-person} kind of consistency, which is not about Bob's
sense data, but rather about facts generally available. A third-person
consistency condition might say that Alice's measurement outcome is a
fact-for-all, and Bob is guaranteed to get the same value if he
performs the same measurement. Previously, Rovelli had endorsed a
premise of
\begin{quote}
  \textbf{Relativity of comparisons:}\ it is meaningless to compare
  the accounts relative to any two systems except by invoking a third
  system relative to which the comparison is made.
\end{quote}
In other words, there is no meaningful comparison between Alice and
Bob, except as seen by Charlie. The revision by Adlam and Rovelli
replaces this first-person orientation with a third-person assumption of
\begin{quote}
  \textbf{Cross-perspective links:}\ In a scenario where some observer
  Alice measures a variable $V$ of a system $S$, then provided that
  Alice does not undergo any interactions which destroy the
  information about $V$ stored in Alice's physical variables, if Bob
  subsequently measures the physical variable representing Alice's
  information about the variable $V$, then Bob's measurement result
  will match Alice's measurement result.
\end{quote}
Up until this point, while the RQM literature endorsed up front a
strong form of relationalism, when one dug into the details, the
writing backed away from it. For example, measurement outcomes were
treated as relative to an observer, but the \emph{choice of which
  measurement was made} would tacitly be treated as a publically
available fact~\cite{DeBrota:2018}. Now, the idea that used to be
implicitly or equivocally stated is brought up front as an
axiom. Indeed, Alice's choice of measurement, her other interactions,
her measurement outcome, Bob's choice of measurement and the result he
obtains are all public, third-person facts in this account. The very
statement of the assumption makes them, in principle,
facts-for-all. If Bob had no way of being aware of whether or not
Alice had disrupted her physical variables, then the assumption would
be a dead letter, entirely inconsequential to his conduct as a
scientist.

So, the assumption \emph{must} matter. But this leads to
difficulties. Take, say, this line from Adlam and Rovelli:
\begin{quote}
[T]he event is an absolute, observer-independent fact, but the value
$v$ is relativized to Alice because at this stage Alice is the only
observer who has this information about $S$, although other observers
could later come to have the same information by interacting
appropriately with either Alice or $S$.
\end{quote}
Is there that big a separation between this view and the one Pauli
expressed in 1958~\cite{Pauli:1994}?
\begin{quote}
Further, personal qualities of the observer do not come into the
theory in any way---the observation can be made by objective
registering apparatus, the results of which are objectively available
for anyone's inspection.
\end{quote}
Or, is there a difference of substance between Adlam and Rovelli's
observers who ``come to have the same information'' and Bohr's
``unambiguously communicable information''~\cite{Bohr:1958}? For Adlam
and Rovelli, those other observers need ``access to Alice and the
ability to perform appropriate measurements''; for Bohr, they need a
``reference to a complete experimental
arrangement''~\cite{Bohr:1958}. How much daylight have the decades brought?

The clearest indication that the original RQM paper was a forward step
beyond old debates was Rovelli's call to reconstruct quantum theory
from physical principles. He made a forceful case that we could see
conceptual progress from technical work in that regard. This naturally
raises a question: What cash value does the introduction of
``cross-perspective links'' have for rederiving the quantum formalism?
To advance beyond where the effort stood in 1996, we have to augment
the principles suggested in the original RQM paper with axioms that
avoid the trap of ``reconstructing'' a theory that can be explained
using local hidden variables. (Local hidden variables can do a lot,
including emulate quantum entanglement; one lesson of Bell and those
that followed is that Schr\"odinger's claim that entanglement held the
essence of quantum theory didn't go far enough~\cite{Bell:1964,
  Werner:1989, Spekkens:2007, Hausmann:2021}.) We could dodge this
trap by forcefully rejecting local hidden variables at the starting
point --- foregrounding the exotic character of quantum
theory~\cite{Stacey:2019, DeBrota:2021}. Or, less dramatically,
mathematical assumptions like the continuity of various important sets
can be invoked, bringing us back eventually to quantum
theory~\cite{Stacey:2021}. Replacing ``relativity of comparisons''
with ``cross-perspective links'' does nothing discernable for the
latter: It is a null move, making those mathematical assumptions
neither more nor less motivated than they were before. As far as the
former approach goes, the replacement makes a first-principles
rejection of hidden variables harder to motivate, since it makes
quantum theory sound \emph{less interesting.} It demands that the
theory turn out ``benignly humdrum''~\cite{Mermin:1989}.

By applying the standard quantum formalism to a system interacting
with an environment, one can show that under the right conditions, the
density matrix of the system will be approximately diagonal in a given
basis. But approximately diagonal is not exactly so; in order to treat
the off-diagonal entries as effectively zero, one needs reason to
regard small numbers as negligible. The standard formalism provides
this via the Born rule, which can take us from small magnitudes in a
density matrix to small probabilities. (Hence why attempts to
\emph{derive} the Born rule from such scenarios turn out circular
sooner or later. One can always try breaking the circle by introducing
new postulates, but if the justification of the project in the first
place was economy of postulates, this hardly seems better.) Moreover,
even a density matrix that is exactly diagonal,
\begin{equation}
  \rho = \begin{pmatrix} p_1 & 0 & 0 & \cdots & 0 \\
    0 & p_2 & 0 & \cdots & 0 \\
    0 & 0 & p_3 & \cdots & 0 \\
    \vdots & \vdots & \vdots & \ddots & \vdots \\
    0 & 0 & 0 & \cdots & p_d
  \end{pmatrix}
\end{equation}
is not the same as a density matrix that is zero everywhere except in
a single entry. Likewise, any calculation within standard quantum
theory that might justify when a ``cross-perspective link'' could be
said to exist will have a statistical character. By manipulating
density matrices, Bob can conclude that he can act as though a
``cross-perspective link'' exists, with some numerical level of
certitude. But this is only a statement about what Bob expects he can
do with low probability of contradiction, not a statement about what
\emph{is.}

The ``cross-perspective link'' postulate tries to promote first-person
data to third-person, but it has no firm grounds on which to say when
that move can be made. We are back to Bell's ``shifty
split''~\cite{Mermin:2018}, to asking how heavy or thermalized an
apparatus must be to qualify as ``classical''~\cite{Camilleri:2015}.

Lahti and Pelonp\"a\"a have recently argued that the
``cross-perspective link'' postulate is ``a disguised form of the
projection postulate leading to a `global collapse' of the state of
the interacting pair''~\cite{Lahti:2022}. This conclusion is difficult
to dodge. In broad terms, if Alice obtains a value $v$ of some
nondegenerate von Neumann observable $V$, then her state for the
observed system is updated to the eigenstate of $V$ corresponding to
that value, $\ket{v}$. But if her value $v$ is a fact-for-all, then
the data establishing the correctness of the state $\ket{v}$ is
factual for all. If ``cross-perspective links'' are consequential,
then $\ket{v}$ should be anybody and everybody's state for that
system. Adlam and Rovelli downplay this, arguing that Bob's state for
the system should reflect his entire history of interactions with it,
and thus its state relative to Bob might not be $\ket{v}$. But this
seems to ignore the weight of the ``cross-perspective links''
assumption. If the fact of Alice obtaining an outcome $v$ is as good
for Bob as obtaining it himself, then it should override his past
history, just as Bob performing a von Neumann measurement himself
would.

\bigskip

This note is based on remarks originally made at
\href{https://scirate.com/arxiv/2203.13342}{SciRate}.

\end{document}